\documentclass[reprint,
 amsmath,amssymb,
aps,
pra,
floatfix,
]{revtex4-1}

\usepackage{graphicx}
\usepackage{dcolumn}
\usepackage{bm}
\usepackage[caption = false]{subfig}
\usepackage{natbib}
\usepackage{url}
\usepackage{hyperref}
\usepackage[mathlines]{lineno}

\begin{document}

\title{Rydberg atoms based creation of N particle GHZ state using STIRAP}


\author{Tanvi P. Gujarati}
\email{tanvipg@umich.edu}
 
\affiliation{%
 Department of Physics, University of Michigan, Ann Arbor, Michigan 48109, USA\\
}%

\date{\today}

\begin{abstract}
Schemes for creation of N particle entangled Greenberger-Horne-Zeilinger (GHZ) states are important for understanding multi-particle non-classical correlations. Here, a theoretical scheme for creation of a multi-particle GHZ state implemented on a target  ensemble of N, $\Lambda$ three-level Rydberg atoms and a single Rydberg atom as a control using Stimulated Raman Adiabatic Passage (STIRAP) is presented. We work in the Rydberg blockade regime for the ensemble atoms induced due to excitation of the control atom to a high lying Rydberg level. It is shown that using STIRAP, atoms from one ground state of the ensemble can be adiabatically transferred to the other ground state, depending on the state of the control atom with high fidelity. Measurement of the control atom in a specific basis after this conditional transfer facilitates one-step creation of a N particle GHZ state. A thorough analysis of adiabatic conditions for this scheme and the influence of radiative decay from the excited Rydberg levels is presented. We show that this scheme is immune to the decay rate of the excited level in ensemble atoms and provides a robust way of creating GHZ states. 

\begin{description}
\item[PACS numbers]
42.50-p, 03.67.-a, 32.80.Rm


\end{description}
\end{abstract}

\maketitle

\section{\label{sec:level1}Introduction}
\begin{figure}
\centering
    \subfloat[Atomic level structure]{\label{fig:Fig1a}\includegraphics[width=0.23\textwidth]{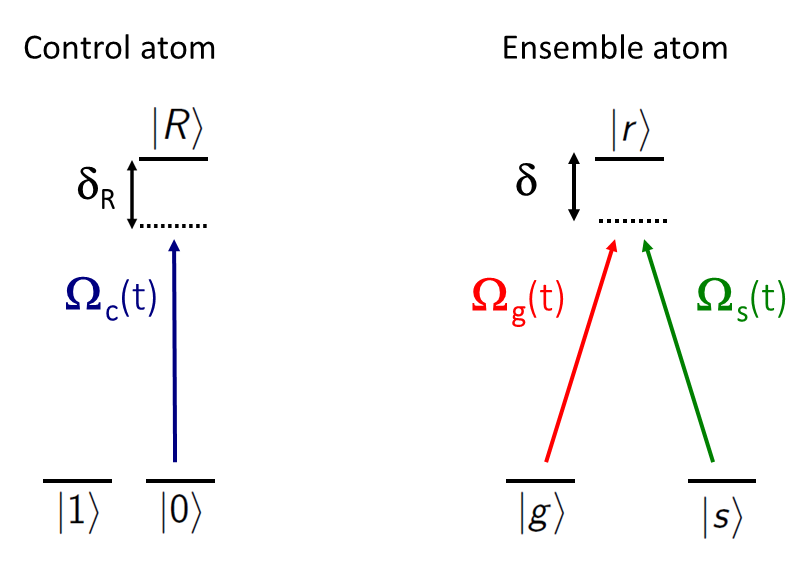}}~              
    \subfloat[Pulse Scheme]{\label{fig:Fig1b}\includegraphics[width=0.23\textwidth]{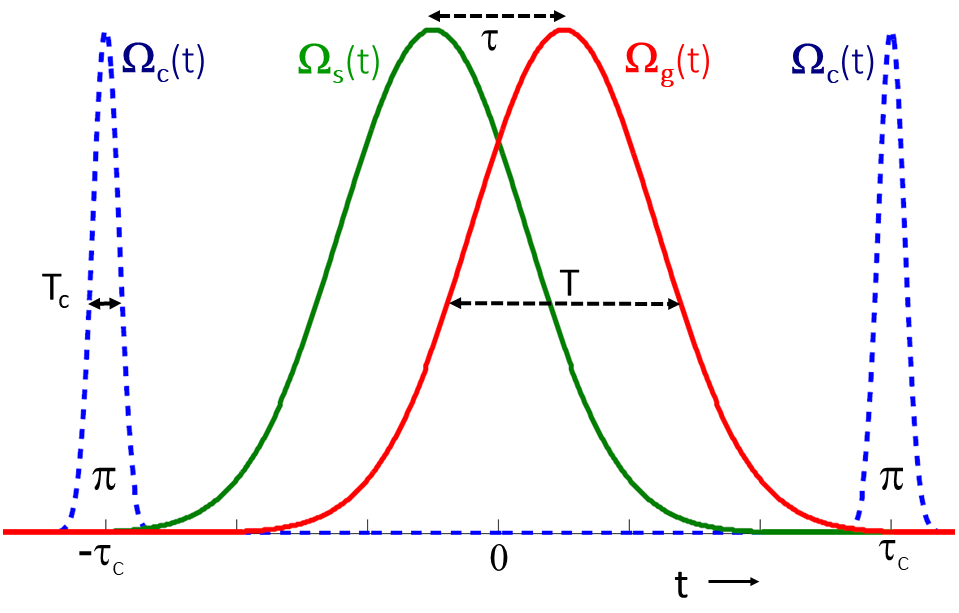}}
\caption{(a)The atomic level structure of the control atom and the target ensemble atoms: The control atom has two metastable states $|0\rangle$ and $|1\rangle$. The level $|0\rangle$ interacts with the excited Rydberg level $|R\rangle$ via Rabi frequency $\Omega_{c}(t)$. $\delta_{R}$ is the detuning between the carrier frequency of the light pulse and the frequency of transition between the levels $|0\rangle$ and $|R\rangle$. The level $|1\rangle$ is isolated from the other levels. Each target atom has a $\Lambda$ type level structure with two metastable states, $|g\rangle$ and $|s\rangle$. They interact with the excited Rydberg level $|r\rangle$ via Gaussian pulses having Rabi frequencies $\Omega_{g}(t)$ and $\Omega_{s}(t)$ respectively. The detuning for both the pulses is given by $\delta$. (b) The pulse sequences: This protocol begins with a Gaussian [$\Omega_{c}(t)$] $\pi$ pulse having a standard deviation given by $T_{c}$ to take the control atom from $|0\rangle$ to $|R\rangle$. It is then followed by counter-intuitive STIRAP pulse sequence with Gaussian profiles, each having $T (\gg T_{c})$ standard deviation. $\tau$ is the time interval between the peaks of these two STIRAP pulses. Finally, another control $\pi$ pulse is used to bring the control atom back to state $|0\rangle$.\label{fig:1}}
\end{figure}

The multi-particle entangled Greenberger-Horne-Zeilinger (GHZ) state shows unique non-local correlations which are essential for understanding the fundamental principles of quantum entanglement \cite{Greenberger89, Carvacho17} and has important applications in quantum information protocols \cite{Zhao04, Kempe99}. Many ingenious schemes for creation of GHZ states in atomic systems have been previously proposed using a multi-step or a single-step process \cite{Muller09, Ostmann17, Saffman10}. In this paper we present a single-step scheme for GHZ state creation employing Rydberg dipole blockade and Stimulated Raman Adiabatic Passage (STIRAP) \cite{Bergmann98, Vitanov17} using a single control atom and an ensemble of target atoms. Rydberg states which are high lying atomic levels, when excited, exert long range dipole forces on the atoms in its vicinity, effectively blocking excitation of more than two atoms to the same Rydberg state \cite{Lukin01,Saffman10}. This phenomenon of `dipole blockade' provides an atomic control that acts on multiple atoms at the same time, which is necessary for generating entanglement between the atoms of the ensemble within the blockade radius.  Approaches to create a multi-particle GHZ state by using Electromagnetically Induced Transparency and adiabatic passage along with Rydberg blockade have been previously studied \cite{Muller09, Unanyan2002, Moller08}. Fidelity of the GHZ states obtained at the end of these protocols is an important parameter to consider. Because of radiative decay from the excited Rydberg states of the ensemble atoms, the fidelity of the GHZ states obtained in these schemes is adversely affected \cite{Saffman10}.\\
Here we propose a different theoretical scheme to realize the creation of a multi-particle GHZ state in an ensemble of $\Lambda$ three-level Rydberg atoms which is robust to radiation decay from the excited Rydberg levels of the ensemble atoms. In this setup, the control atom and the ensemble of the target atoms are assumed to be independently addressable. This can be achieved by storing them in two separate trapping potentials in close proximity or in a lattice where the control atom can be efficiently addressed. This setup is similar to what has been discussed in the proposal by Muller et. al. \cite{Muller09}. \\
The control atom has a three level structure as is shown in Fig. (\ref{fig:Fig1a}). The two metastable levels $|0\rangle$ and $|1\rangle$ determine the state of the control atom. Level $|0\rangle$ is connected to the excited Rydberg level $|R\rangle$ via a control pulse with Rabi frequency given by $\Omega_{c}(t)$.


Level $|1\rangle$ is chosen such that dipolar transitions between $|1\rangle$ and $|0\rangle$ as well as $|R\rangle$ are forbidden. An ensemble having N target Rydberg atoms is considered to be within the blockade radius of the excited control atom. The level structure of the ensemble atoms and the corresponding pulse sequence acting on them is shown in Fig. (\ref{fig:1}). Every ensemble atom has two metastable ground states, namely, $|g\rangle$ and $|s\rangle$ and one Rydberg excited level $|r\rangle$. All the ensemble atoms are initiated in the $|g\rangle$ state.  This GHZ state creation protocol begins with a control $\pi$ pulse having Rabi frequency $\Omega_{c}(t)$ which is used to excite the control atom. If the control atom is in state $|1\rangle$, the control pulse has no effect. On the other hand, if it is in state $|0\rangle$, with the action of the control $\pi$ pulse, the atom is excited to the Rydberg level $|R\rangle$. Due to the long range dipole-dipole interactions between the excited Rydberg level $|R\rangle$ and Rydberg levels $|r\rangle$, the target ensemble Rydberg levels undergo energy level shift given by a frequency $\Delta$. In the absence of this energy shift, the condition for adiabatic population transfer of the ensemble atoms from the ground state $|g^{N}\rangle=\otimes_{j=1}^{N}|g\rangle_{j}$ to $|s^{N}\rangle=\otimes_{j=1}^{N}|s\rangle_{j}$ via the counter-intuitive STIRAP pulse sequence $\Omega_{s}(t)$ and $\Omega_{g}(t)$ [Fig. (\ref{fig:1})] is satisfied. The parameters of the system are set up in such a way that when the control atom is excited to $|R\rangle$, the induced energy shift $\Delta$ in the ensemble atoms disrupts the STIRAP condition for population transfer from $|g^{N}\rangle$ to $|s^{N}\rangle$. Due to the added detuning the population remains in the state $|g^{N}\rangle$ after the application of the STIRAP pulses. Finally, another control $\pi$ pulse is used to bring the control atom back to the original state. When the control atom is prepared in the $\frac{1}{\sqrt{2}}(|0\rangle + |1\rangle)$ superposition state at the beginning of the protocol and finally measured in the superposition basis, the ensemble atoms get projected to a N particle GHZ state.\\
If the conditions for STIRAP are met, the instantaneous eigenstate occupied by the ensemble atoms has no contribution from the level $|r\rangle$ at all times. Hence, this protocol is insensitive to the radiative decay losses from the excited Rydberg level of the ensemble atoms. \\
Let us now analyze this scheme in detail and study the dependence of the STIRAP transfer conditions on the parameters of the system. In Sec. \ref{sec:level2} we discuss the dynamics of the control atom. This is followed by the discussion of the transfer mechanism in the target atoms and the adiabaticity conditions required for efficient transfer in Sec. \ref{sec:level3}. Numerical simulations of this protocol for realistic parameters are then presented in Sec. \ref{sec:level4} . In Sec. \ref{sec:level5} we conclude the discussion.

\section{\label{sec:level2}The control atom}
\begin{figure}
\centering
    \subfloat[$T_{c} = 0.1 T$]{\label{fig:Fig2a}\includegraphics[width=0.24\textwidth]{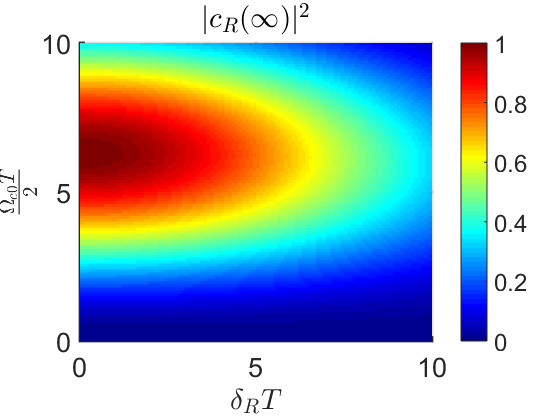}}~              
    \subfloat[$T_{c} = 1 T$]{\label{fig:Fig2b}\includegraphics[width=0.24\textwidth]{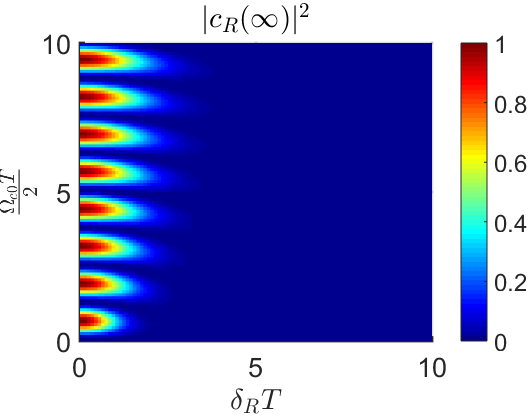}}
\caption{(a) The coefficient of population in state $|R\rangle$ transferred from $|0\rangle$, $|c_{R}(\infty)|^{2}$, due to the control $\pi$ pulse is plotted as a function of scaled detuning $\delta_{R}T$ and scaled peak control Rabi frequency $\Omega_{c0}T$ for a value of $T_{c} = 0.1 T$. (b) Same as plot (a) but for value of $T_{c}=1 T$. We see that smaller values of $T_{c}$ are more robust to variations in detuning and peak Rabi frequency.}\label{fig:2} 
\end{figure}

Hamiltonian for the control atom interacting with the classical control field in the field interaction representation with the rotating wave approximation (RWA) is given below:
\begin{eqnarray}
\frac{H_{C}(t)}{\hbar}&=& \delta_{R}|R\rangle\langle R|+\frac{\Omega^{*}_{c}(t)}{2}|0\rangle\langle R|+\frac{\Omega_{c}(t)}{2}|R\rangle\langle 0|
\label{eq:conH}
\end{eqnarray}
The energy levels are measured relative to the ground state energy $\hbar\omega_{0}=0$. In Eq. (\ref{eq:conH}), $\delta_{R}\equiv\omega_{R}-\omega_{c}$ is the detuning between the frequency of transition from $|0\rangle$ to $|R\rangle$ (denoted by $\omega_{R}$) and the optical frequency of the control pulse, $\omega_{c}$. As noted previously, $\Omega_{c}(t)$ is the Rabi frequency of the control pulse with a Gaussian temporal profile given below.
\begin{eqnarray}
\Omega_{c}(t)&=&\Omega_{c0}\exp\big[{-\frac{(t-\tau_{c})^2}{2T_{c}^{2}}\big]}
\end{eqnarray}
We will assume $\Omega_{c0}$ to be real in all the calculations here after. The level $|1\rangle$ is isolated from the levels $|0\rangle$ and $|R\rangle$. For $\delta_{R}=0$, on solving the Schrodinger's equation for a general wave-function, $|\Psi(t)\rangle = c_{0}(t)|0\rangle+c_{R}(t)|R\rangle$, with $|c_{0}(-\infty)| = 1$, we get:
\begin{eqnarray}
|c_{0}(\infty)|^{2}&=&\cos^{2}\Theta\\
|c_{R}(\infty)|^{2}&=&\sin^{2}\Theta\\
\Theta \equiv \int_{-\infty}^{\infty}\frac{\Omega_{c}(t')}{2}dt' &=& \Omega_{c0}T_{c}\sqrt{\frac{\pi}{2}}
\end{eqnarray}
For complete transfer of population from $|0\rangle$ to $|R\rangle$ state, $\Theta$ should be an odd multiple of $\frac{\pi}{2}$. Thus, we need:\\
\begin{eqnarray}
\Omega_{c0}T_{c}&=& (2p+1)\sqrt{\frac{\pi}{2}},~~~~p\in\mathbb{Z}
\label{eq:oddmul}
\end{eqnarray}
To check for the robustness of this transfer against variations in the Rabi frequency, we look at the derivative of $|c_{R}(\infty)|$ wrt $\Omega_{c0}$.
\begin{eqnarray}
\frac{\partial|c_{R}(\infty)|}{\partial\Omega_{c0}}&=&-T_{c}\sqrt{\frac{\pi}{2}}\cos(\Omega_{c0}T_{c}\sqrt{\frac{\pi}{2}})
\label{eq:robust}
\end{eqnarray}
Eq.~(\ref{eq:robust}) implies that smaller values of $T_{c}$ provide more robustness against variation in $\Omega_{c0}$. For $\delta_{R}\neq 0$, analytic solution for Gaussian form of the Rabi frequency is difficult to derive. Hence, we will look at the dependence of $|c_{R}(\infty)|^{2}$ on different values of $\Omega_{c0}$, $\delta_{R}$ and $T_{c}$ numerically in Fig. (\ref{fig:2}). For the value of $T_{c} = 0.1T$, where $T$ is the standard deviation of the Gaussian STIRAP pulses, we see from Fig. (\ref{fig:Fig2a}) that the population gets completely transferred to the $|R\rangle$ state when $\Omega_{c0}T = 6.2$ and $\delta_{R}T = 0$. From Fig. (\ref{fig:Fig2b}), we see that there are multiple periodic values of $\Omega_{c0}T$ for which complete population transfer to the excited level can be achieved via a $\pi$ pulse as expected from Eq. (\ref{eq:oddmul}) for $T_{c}=1T$. As $\delta_{R}T$ becomes larger, the fraction of population in the excited state decreases and eventually becomes zero. The effect of larger values of $\delta_{R}T$ is more prominent for larger  values of $T_{c}$. As derived in Eq. (\ref{eq:robust}), we see that smaller values of $T_{c}$ provide more robust transfer against variations in $\Omega_{c0}$ and $\delta_{R}$.  

\section{\label{sec:level3}The target ensemble}
In this section, we will derive the conditions that are necessary to maintain adiabatic transfer of the ensemble atoms from $|g^{N}\rangle$ to $|s^{N}\rangle$ when the control atom is in state $|1\rangle$ and to remain in the state $|g^{N}\rangle$ when the control atom is in the $|0\rangle$ state. The Hamiltonian for ensemble atoms interacting with the counter-intuitive STIRAP pulse sequence in the RWA is given below: 
\begin{eqnarray}
\frac{H_{T}(t)}{\hbar}&=& \sum_{j=1}^{N}\big[(\omega_{r}^{0}-\delta_{g})|g\rangle_{j}\langle g|+(\omega_{r}^{0}-\delta_{s})|s\rangle_{j}\langle s|\big]\nonumber\\
&+&\sum_{j=1}^{N}\big[\frac{\Omega^{*}_{g}(t)}{2}e^{-\textit{i}\omega^{0}_{r}t}|g\rangle_{j}\langle r|+\frac{\Omega^{*}_{s}(t)}{2}e^{-\textit{i}\omega^{0}_{r}t}|s\rangle_{j}\langle r| \nonumber\\
&&+\text{h.c.}\big]
\label{eq:tarH}
\end{eqnarray}
In Eq. (\ref{eq:tarH}), $\hbar\omega_{r}^{0}$ is the energy of the excited level $|r\rangle$. For the energy of states $|g\rangle$ and $|s\rangle$ denoted by $\hbar\omega_{g}^{0}$ and $\hbar\omega_{s}^{0}$ respectively,  $\delta_{g(s)} = \omega_{r}^{0}-\omega_{g(s)}^{0}-\omega_{g(s)}$ are the detunings of these levels wrt to the optical frequencies $\omega_{g}$ and $\omega_{s}$ of the STIRAP pulses shown in Fig. (\ref{fig:Fig1b}). The corresponding Rabi frequencies $\Omega_{g}(t)$ and $\Omega_{s}(t)$ are defined as follows:
\begin{eqnarray}
\Omega_{g}(t)&=&\Omega\exp{\big[-\frac{(t-\frac{\tau}{2})^{2}}{2T^{2}}\big]}\label{eq:omg}\\
\Omega_{s}(t)&=&\Omega\exp{\big[-\frac{(t+\frac{\tau}{2})^{2}}{2T^{2}}\big]}
\label{eq:oms}
\end{eqnarray}
In Eqs. (\ref{eq:omg})-(\ref{eq:oms}), $\Omega$ is the peak Rabi frequency of the Gaussian STIRAP pulses, $\tau$ is the time separation between the peaks of the two pulses and $T$ is the standard deviation. We can simplify the Hamiltonian in Eq. (\ref{eq:tarH}) by setting $\omega_{r}^{0}=0$ and assuming two photon resonance condition for the system i.e. $\delta_{g}=\delta_{s}=\delta$ \cite{Bergmann98}. Boosting the energy of all the levels by $\delta$, we get the modified Hamiltonian for the target ensemble as: 
\begin{eqnarray}
\frac{H_{T}(t)}{\hbar}&=& \sum_{j=1}^{N}\big[\frac{\delta}{2}|r\rangle_{j}\langle r|+\frac{\Omega^{*}_{g}(t)}{2}|g\rangle_{j}\langle r|+\frac{\Omega^{*}_{s}(t)}{2}|s\rangle_{j}\langle r| \nonumber\\
&&+\text{h.c.}\big]
\label{eq:tarH2}
\end{eqnarray}
 We will restrict the set of basis states for the analysis of this system to a set containing only one Rydberg level excitation by assuming that all the atoms are within the Rydberg blockade radius of each other. We can rewrite the Hamiltonian in Eq. (\ref{eq:tarH2}) in the symmetric Fock state basis set defined by:
\begin{eqnarray}
\Sigma_{\mu,\nu}&=&\sum_{j}|\mu\rangle_{j}\langle \nu| = a_{\mu}^{\dagger}a_{\nu};\\
|g^{N-n};s^{n};r^{0}\rangle& =& \sqrt{\frac{(N-n)!}{N!n!}}\Sigma_{s,g}^{n}|g^{N}\rangle\\
|g^{N-n-1};s^{n};r^{1}\rangle &=& \sqrt{\frac{(N-n-1)!}{N!n!}}\Sigma_{s,g}^{n}\Sigma_{r,g}|g^{N}\rangle
\end{eqnarray}
There are in all (2N+1) states in this basis set, namely,
\begin{eqnarray}
\big\{|g;s;r\rangle_{N}\big\}&=&\big\{|g^{N};s^{0};r^{0}\rangle,..,|g^{N-n};s^{n};r^{0}\rangle,..|g^{0};s^{N};r^{0}\rangle, \nonumber\\
& &|g^{N-1};s^{0};r^{1}\rangle,..,|g^{N-n-1};s^{n};r^{1}\rangle,..|g^{0};s^{N-1};r^{1}\rangle\big\} \nonumber\\
\label{eq:set}
\end{eqnarray}
As a short hand notation, we use $|g^{N}\rangle \equiv |g^{N};s^{0};r^{0}\rangle$ and $|s^{N}\rangle \equiv |g^{0};s^{N};r^{0}\rangle $. The corresponding Hamiltonian in the Fock number basis is then:
\begin{eqnarray}
\frac{H_{T}(t)}{\hbar}&=&\delta \sigma_{r}^{+}\sigma_{r}^{-}+\big[\frac{\Omega_{g}^{*}(t)}{2}a_{g}^{\dagger}\sigma_{r}^{-}+\frac{\Omega_{s}^{*}(t)}{2}a_{s}^{\dagger}\sigma_{r}^{-}+\text{h.c.}\big]\nonumber\\
\label{eq:tarH3}
\end{eqnarray}
Where:
\begin{eqnarray}
\sigma_{r}^{+}|r^{0}\rangle &=&  |r^{1}\rangle ~~~~~~\sigma_{r}^{-}|r^{0}\rangle = 0\\
\sigma_{r}^{-}|r^{1}\rangle &=& |r^{0}\rangle ~~~~~~\sigma_{r}^{+}|r^{1}\rangle =  0
\end{eqnarray}
Using the properties of block tri-diagonal matrices it can be shown that the Hamiltonian in Eq. (\ref{eq:tarH3}) when expressed as a matrix in the basis set defined by Eq. (\ref{eq:set}) always has one eigenvalue as 0. The characteristic equation for this Hamiltonian is invariant when $\delta \rightarrow -\delta$ and the eigenvalue $\lambda \rightarrow -\lambda$. This structure implies that the other 2N eigenvalues are symmetrically placed around the eigenvalue 0. With the following new definitions given in Eqs. (\ref{eq:om0})-(\ref{eq:the}), we are set to explore the eigen-structure of this system.
\begin{eqnarray}
\Omega_{0}(t) &\equiv& \sqrt{\Omega_{g}^{2}(t)+\Omega_{s}^{2}(t)} \label{eq:om0}\\
\tan\theta(t) \equiv \frac{\Omega_{g}(t)}{\Omega_{s}(t)};&~~~&\tan\varphi(t) \equiv \frac{\Omega_{0}(t)}{\delta}\label{eq:the}
\end{eqnarray}
\begin{figure}
\centering
    \subfloat[$|c_{s}(\infty)|^{2}$]{\label{fig:Fig3a}\includegraphics[width=0.25\textwidth]{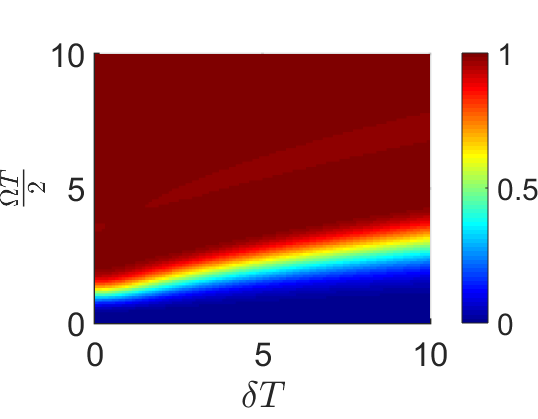}}~              
    \subfloat[$|c_{s}(\infty)|^{2}+|c_{g}(\infty)|^{2}$]{\label{fig:Fig3b}\includegraphics[width=0.25\textwidth]{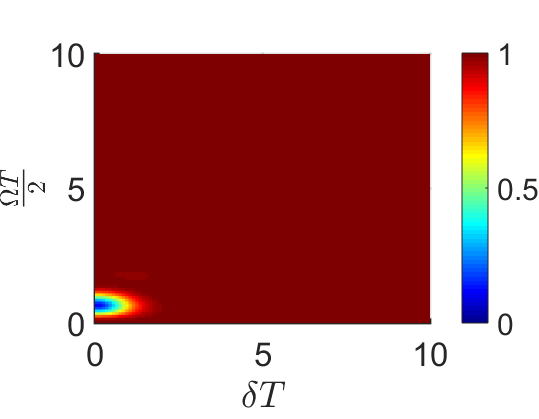}}\\
    \subfloat[$|c_{s^{5}}(\infty)|^{2}$]{\label{fig:Fig3c}\includegraphics[width=0.25\textwidth]{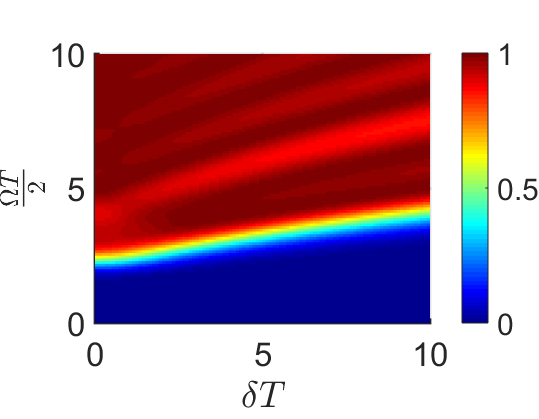}}~              
    \subfloat[$|c_{s^{5}}(\infty)|^{2}+|c_{g^{5}}(\infty)|^{2}$]{\label{fig:Fig3d}\includegraphics[width=0.25\textwidth]{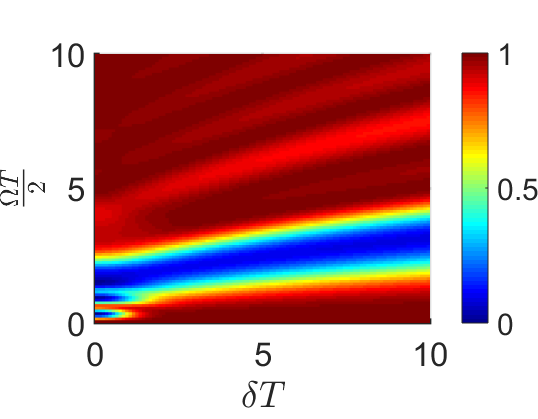}}\\
\caption{(a) Co-efficient of population in state $|s\rangle$ for a target ensemble with 1 atom after the application of STIRAP pulses as a function of the scaled peak Rabi frequency $\Omega T$ and scaled detuning $\delta T$ for $\tau = 1.4T$.(b)Total population in the state $|s\rangle$ and $|g\rangle$ after the STIRAP pulses for a single target atom as a function of $\Omega T$ and $\delta T$. (c) Same as plot (a) but for a target ensemble of 5 atoms. (d) Same as plot (b) for N = 5 atoms. We see that as the number of target atoms goes up, the parameter space for adiabatic transfer from $|g^{N}\rangle$ to $|s^{N}\rangle$ or no transfer gets modified as per the conditions derived in Eqs. (\ref{eq:a})-(\ref{eq:c}) }\label{fig:3}
\end{figure}
On solving for the eigenvalues of this system, we find that the non-zero eigenenergies are:
\begin{eqnarray}
E^{N}_{\pm n} &=& \frac{\hbar\Omega_{0}(t)}{2}[\cot\varphi(t)\pm\sqrt{n+\cot^{2}\varphi(t)}], ~~~~ n = 1,..,N\nonumber\\
\end{eqnarray}
The corresponding eigenstates are be denoted by $|\lambda_{\pm n}^{N}\rangle$. The eigenstate with eigenenergy 0 is given as:
\begin{eqnarray}
|O(t)\rangle &=& \sum_{n=0}^{N}(-1)^{N-n}\alpha^{N}_{n}(t)|g^{N-n};s^{n};r^{0}\rangle\\
\alpha^{N}_{n}(t)&=&\sqrt{\frac{N!}{n!(N-n)!}}\cos^{N-n}(\theta(t))\sin^{n}(\theta(t))
\end{eqnarray}
State $|O(t)\rangle$ is the N particle STIRAP state. As $t\rightarrow -\infty$, $|O(-\infty)\rangle = |g^{N}\rangle$ and $t\rightarrow \infty$, $|O(\infty)\rangle = |s^{N}\rangle$. If this system evolves adiabatically, then the population of the target ensemble can be coherently transferred from $|g^{N}\rangle$ to $|s^{N}\rangle$. This eigenstate with eigenvalue 0 has no contribution from the excited level $|r\rangle$ for any number of ensemble atoms at all times. It is also independent of the detuning $\delta$. In the STIRAP process our aim is to keep the target ensemble in the instantaneous eigenstate $|O(t)\rangle$ at all times. Adiabatic population transfer along this eigenstate implies that this protocol is insensitive to the spontaneous emissions from the excited level $|r\rangle$. This is a key feature of this scheme which provides us with a robust mechanism of population transfer even in the presence of decay. Numerical studies in the presence of decay are described in Sec. \ref{sec:level4}. \\
The condition for maintaining adiabatic transfer along the $|O(t)\rangle$ state is summarized by the adiabaticity criterion discussed in \cite{Comparat09} given as:
\begin{eqnarray}
\sum_{m\neq 0}\big|\frac{\hbar\langle m|\dot {O}(t)\rangle}{E_{0}-E_{m}}\big|\ll 1
\label{eq:adcond}
\end{eqnarray}
In the above Eq. (\ref{eq:adcond}), $E_{0}$ is the eigenenergy of the eigenstate $|O(t)\rangle$ and the sum is taken over all the other eigenstates $|m\rangle$ with eigenenergies $E_{m}$.\\
From here onwards, we will assume $\Omega$ to be real. On analyzing the eigenstates $|\lambda^{N}_{\pm 1}\rangle$ corresponding to eigenenergies $E^{N}_{\pm 1}$, we find that the projection of state $|\lambda^{N}_{\pm 1}\rangle$ onto the $|r^{0}\rangle$ subspace is co-linear with $|\dot{O}(t)\rangle$:

\begin{eqnarray}
\langle \lambda^{N}_{+ 1}(t)|\dot{O}(t)\rangle &=& \frac{\dot{\theta}(t)\sqrt{N}\sin(\frac{\varphi(t)}{2})}{\cot(\frac{\varphi(t)}{2})}\\ 
\langle \lambda^{N}_{- 1}(t)|\dot{O}(t)\rangle &=& \frac{-\dot{\theta}(t)\sqrt{N}\cos(\frac{\varphi(t)}{2})}{\tan(\frac{\varphi(t)}{2})}
\end{eqnarray}
The eigen-structure is such that for any value of N, all the eigenstates except the zeroth eigenstate have non-zero projections in the $|r^{1}\rangle$ subspace. From the orthonormality properties of the eigenvectors we can deduce that:
\begin{eqnarray}
\langle \lambda^{N}_{\pm n}|P_{r^{0}}P^{\dagger}_{r^{0}}|\lambda^{N}_{\pm m}\rangle = \langle \lambda^{N}_{\pm n}|P_{r^{1}}P^{\dagger}_{r^{1}}|\lambda^{N}_{\pm m}\rangle = 0 ~~\forall~~n\neq m\nonumber\\
\end{eqnarray}
Here, $P^{\dagger}_{r^{0}}$ and $P^{\dagger}_{r^{1}}$ are projection operators for the $|r^{0}\rangle$ and $|r^{1}\rangle$ subspace respectively. From the above deduction we can conclude that only the $|\lambda^{N}_{\pm 1}\rangle$ eigenstates contribute to the sum in Eq. (\ref{eq:adcond}). On simplifying the adiabatic condition we get:
\begin{eqnarray}
\dot{\theta}(t) &\ll&\frac{\Omega_{0}(t)}{2\sqrt{N}}f(\varphi(t))\label{eq:thetdot}\\
f(\varphi(t))&=&\frac{\sin\frac{\varphi(t)}{2}\cos\frac{\varphi(t)}{2}}{\sin^{3}\frac{\varphi(t)}{2}+\cos^{3}\frac{\varphi(t)}{2}}\label{eq:adexp}
\end{eqnarray}
Substituting the expressions for $\Omega_{0}(t)$ and $\dot{\theta}(t)$ in Eq. (\ref{eq:thetdot}), the adiabaticity condition is rewritten in Eq. (\ref{eq:adexp2}). Here, we have scaled all the variables with $T$, thus, $\tilde{\Omega}\equiv\Omega T$, $\tilde{\tau}\equiv\frac{\tau}{T}$ and similarly $\tilde{\delta}$ and $\tilde{t}$.
\begin{eqnarray}
1&\ll&\sqrt{\frac{2}{N}}\frac{\tilde{\Omega}}{\tilde{\tau}}  \exp{(-\frac{(\tilde{t}^{2}+\frac{\tilde{\tau}^{2}}{4})}{2})}\cosh^{3/2}(\tilde{t}\tilde{\tau})f(\varphi(\tilde{t}))
\label{eq:adexp2}
\end{eqnarray}
Since, the Rabi frequencies and detuning are positive, $0 \leq \varphi(t) < \frac{\pi}{2}$. The function $f(\varphi(t))$ is a monotonically increasing function of $\varphi(t)$ in this range. For the strictest adiabaticity condition, we should choose the limit when $\varphi(t)\rightarrow 0 $. In this limit, $f(\varphi(t)) = \frac{\Omega_{0}(t)}{\sqrt{2}\delta}$, given $\delta \gg \Omega_{0}(t)$. On the other hand, when $\varphi(t) \rightarrow \frac{\pi}{2}$, we get $f(\varphi(t)) = \frac{1}{\sqrt{2}}$ with $\delta \rightarrow 0$. For the duration of population transfer, i.e. when $\Omega_{0}(\tilde{t})$ is considerably large, the $\tilde{t}$ dependence of the RHS of Eq. (\ref{eq:adexp2}) varies from being singly peaked with maximum at $\tilde{t}=0$ till $\tilde{\tau}$ is increased from 0 to about 1.4 to being doubly peaked as $\tilde{\tau}$ is increased further with a minimum at $\tilde{t}=0$. It is thus sufficient to study the Eq. (\ref{eq:adexp2}) at $\tilde{t}=0$ for all values of $\tilde{\tau}$. Incorporating the above simplifications, the adiabaticity condition now is given as:
\begin{eqnarray}
1&\ll&\frac{\tilde{\Omega}^2}{\sqrt{N}\tilde{\tau}\tilde{\delta}} \exp{(-\frac{\tilde{\tau}^{2}}{4})}~~~~\text{when}~~~\tilde{\delta}\gg\tilde{\Omega}
\label{eq:adconddel}
\end{eqnarray}
It is worthwhile to keep in mind that when $\delta \rightarrow 0$, this condition becomes:
\begin{eqnarray}
1&\ll&\frac{\tilde{\Omega}}{\sqrt{N}\tilde{\tau}} \exp{(-\frac{\tilde{\tau}^{2}}{8})}
\label{eq:adcondnodel}
\end{eqnarray}
Note the dependence of the adiabaticity conditions in Eq. (\ref{eq:adconddel}) and Eq. (\ref{eq:adcondnodel}) on the number of atoms in the ensemble. The condition for adiabatic transfer along the $|O\rangle$ eigenstate becomes stricter by $\sqrt{\text{N}}$ for an ensemble of N atoms. The optimum value of $\tau$ can be obtained numerically. When all other parameters are fixed, the condition $\tilde{\delta} \ll \tilde{\Omega}^{2}$ for the adiabatic transfer is similar to what was proved by Vitanov and Stenholm in 1997 \cite{Vitanov97} for a single atom case.\\

Let us now understand the condition required for the atomic population to remain in the state $|g^{N}\rangle$ when the added detuning due to Rydberg dipole-dipole interaction is introduced. For a single atom case, as long as $\tilde{\delta}\gg\tilde{\Omega}$, we can reduce the three level system to a two level system. In this case, the condition for adiabatic transfer from $|g\rangle$ to $|s\rangle$ is simply $\tilde{\delta}\ll\tilde{\Omega}^{2}$, ignoring the effects of $\tilde{\tau}$. On the other hand, the condition to remain in the $|g\rangle$ state is $\tilde{\Omega}^2 \ll \tilde{\delta}$ which is obtained by making the effective coupling between levels $|g\rangle$ and $|s\rangle$ small \cite{Vitanov97}. This situation changes a little in the presence of more than one atom. In this case, when we enforce that the effective couplings are kept small, the condition for the ensemble state to remain in the state $|g^{N}\rangle$ is modified to:
\begin{eqnarray}
\sqrt{N}\tilde{\Omega}^{2} \ll \tilde{\delta}~~~\text{when}~~~\tilde{\Omega}\ll\tilde{\delta} 
\end{eqnarray}
Thus, we can conclude that for the ensemble state to be transferred to $|s^{N}\rangle$ state from the initial state $|g^{N}\rangle$, assuming $\tilde{\tau}$ is fixed, we must have:
\begin{eqnarray}
\delta_{|1\rangle} &\ll& \frac{\tilde{\Omega}^{2}}{\sqrt{N}} ~~~\text{when}~~~\tilde{\delta}_{|1\rangle}\gg\tilde{\Omega}
\label{eq:a}\\
1 &\ll& \frac{\tilde{\Omega}}{\sqrt{N}} ~~~\text{when}~~~\tilde{\delta}_{|1\rangle}\rightarrow 0
\label{eq:b}
\end{eqnarray}
Also, for the ensemble state to remain in the $|g^{N}\rangle$ state, we must have:
\begin{eqnarray}
\delta_{|0\rangle} &\gg&\sqrt{N}\tilde{\Omega}^{2} ~~~\text{when}~~~\tilde{\delta}_{|0\rangle}\gg\tilde{\Omega}
\label{eq:c}
\end{eqnarray}
In the above equations $\tilde{\delta}_{|0\rangle}$ and $\tilde{\delta}_{|1\rangle}$ are the detunings of ensemble atoms when the control atom is in state $|0\rangle$ and $|1\rangle$ respectively. For our protocol to work efficiently, our system should satisfy the conditions given in Eq. (\ref{eq:a}) or Eq. (\ref{eq:b}) along with Eq. (\ref{eq:c}). Thus, we can take $\tilde{\delta}_{|0\rangle} = \tilde{\delta}_{|1\rangle} + \tilde{\Delta}$. \\
To understand the implications of the adiabaticity conditions derived in this section, we numerically evolve the Hamiltonian for the ensemble atoms given in Eq. (\ref{eq:tarH3}) for different values of $\Omega T$ and $\delta T$. In Fig. (\ref{fig:3}) we plot the population of ensemble atoms in state $|s^{N}\rangle$ for $N = 1$ and $5$ denoted by the co-efficient $|c_{s^{N}}(\infty)|^2$. To compare this with the population that remained in the initial state $|g^{N}\rangle$, we plot the total population in the states $|s^{N}\rangle$ and $|g^{N}\rangle$ after the completion of the protocol. This sum is denoted as $|c_{g^{N}}(\infty)|^{2}+|c_{s^{N}}(\infty)|^{2}$. For $N=1$, we see from Fig. (\ref{fig:Fig3a}), the population gets completely transferred to $|s\rangle$ state for $\tilde{\Omega}^{2}\gg\tilde{\delta}$. It is clear from Fig. (\ref{fig:Fig3b}), there is only a small portion of the parameter space when $\tilde{\Omega}\approx\tilde{\delta} < 3$ where adiabatic transfer of population as described above does not take place for $N=1$. This situation changes as the number of atoms in the target ensemble increases since more intermediate states now become available. For $N=5$, as seen from Fig. (\ref{fig:Fig3c}), the condition for adiabatic transfer from $|g^{5}\rangle$ to $|s^{5}\rangle$ becomes stricter compared to that for $N=1$. Portions of the parameter space defined by $\tilde{\Omega}$ and $\tilde{\delta}$ open up where the adiabaticity conditions fail. This region clearly divides the parameter space into two sections, one which allows the adiabatic transfer of population from $|g^{N}\rangle$ to $|s^{N}\rangle$ with high fidelity marked out by the condition $\tilde{\Omega}^{2}\gg\sqrt{N}\tilde{\delta}$ and the other where population remains in $|g^{N}\rangle$ with unit probability. The Rydberg-Rydberg interaction between the control and the ensemble atoms provides a tunable mechanism to increase or decrease the effective value of $\tilde{\delta}$ such that the target atoms are always in either of these two high fidelity transfer regions subject to the state of the control atom. 

\section{\label{sec:level4}Numerical results}
Before we start analyzing the numerical simulations for the control and target system together, let us introduce the effect of decoherence due to spontaneous emissions from the excited Rydberg states for the control atom and the target ensemble. \\
Assuming no collisions, the master equation for the density matrix, $\rho$, with $M$ number of spontaneous emission decay channels is given below:
\begin{eqnarray}
\dot{\rho}&=&\frac{\textit{i}}{\hbar}[\rho,H]+\hat{L}(\rho)\\
\hat{L}(\rho)&=&-\frac{1}{2}\sum_{m=1}^{M}(C_{m}^{\dagger}C_{m}\rho\text{+}\rho C_{m}^{\dagger}C_{m})\text{+}\sum_{m=1}^{M}C_{m}\rho C^{\dagger}_{m}
\end{eqnarray}
For the control atom, we have only one decay channel with the decay rate $\Gamma_{0R}$, namely,
\begin{eqnarray}
\hat{C}_{0R}&=&\sqrt{\Gamma_{0R}}|0\rangle\langle R|
\end{eqnarray}
For the target ensemble atoms, there are two decay channels with rates $\Gamma_{gr}$ and $\Gamma_{sr}$ defined as:
\begin{eqnarray}
\hat{C}_{gr}&=&\sqrt{\Gamma_{gr}}|g\rangle\langle r|\\
\hat{C}_{sr}&=&\sqrt{\Gamma_{sr}}|s\rangle\langle r|
\end{eqnarray}
In the forth coming numerical calculations, we have chosen $\Gamma_{gr}=\Gamma_{sr}\equiv\Gamma_{r}$. It is straight-forward to extend the master equation calculations for a system with more than one target atom using the Fock number state basis. \\
\begin{figure}
\centering
    \includegraphics[width=0.4\textwidth]{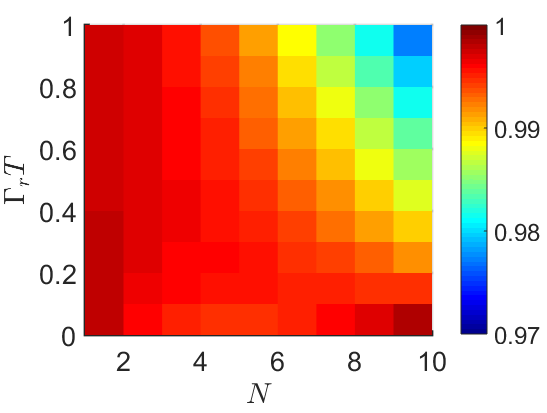}
\caption{The population in level $|s^{N}\rangle$ after the STIRAP pulses for different number of atoms in the target ensemble, N, and varying spontaneous emission rate $\Gamma_{r}T$. The value of detuning $\delta T =0$, $\Omega T = 9.5$ and $\tau = 1.4T$. We see that the population transfer does not depend on the decay rate significantly and has values higher than 0.99 for typical range of $\Gamma_{r}T\approx 0.01\sim0.1$ }\label{fig:4}
\end{figure}
We will first study the effect of decay due to spontaneous emissions on the target ensemble with different number of atoms. We choose the value of $T=1\mu s$ and $\tilde{\tau} = 1.4$ for all the numerical results here after. From Fig. (\ref{fig:4}), we see that even for an ensemble of about ten atoms, the population transferred to the $|s^{N}\rangle$ state from the $|g^{N}\rangle$ state is greater than 99\% for realistic values of Rydberg level spontaneous emission rates of about $\Gamma_{r}\approx 0.01\sim 0.1$ MHz. As discussed above, we see that the spontaneous emissions from the Rydberg excited levels of the target ensemble atoms do not affect this protocol which makes it a very robust scheme. \\
Having laid the groundwork we will now look at the simulation of GHZ state creation. The total Hamiltonian for this system is:
\begin{eqnarray}
H_{Tot}(t)&=&H_{C}(t)+H_{T}(t)+\hbar\Delta |R\rangle \sigma_{r}^{+}\sigma_{r}^{-}\langle R |
\label{eq:totH}
\end{eqnarray}
The expressions for $H_{C}(t)$ and $H_{T}(t)$ are given in the Eq. (\ref{eq:conH}) and Eq. (\ref{eq:tarH3}) respectively. The interaction between the target ensemble and the control atom is introduced via the last term in Eq. (\ref{eq:totH}) with the interaction strength given by frequency $\Delta$.\\
We solve the Schrodinger equation numerically in the basis set $\big\{|0\rangle,|1\rangle,|R\rangle\big\}\otimes\big\{|g;s;r\rangle_{N}\big\}$ defined in Eq. (\ref{eq:set}) with the Hamiltonian defined by Eq. (\ref{eq:totH}) for the control atom in the initial state, $\frac{1}{\sqrt{2}}(|0\rangle+|1\rangle)$ and the ensemble atoms initiated in the $|g^{N}\rangle$ state. In Fig. (\ref{fig:5}) we have plotted the modulus squared of the co-efficients corresponding to the components $|0\rangle|g^{N}\rangle$, $|0\rangle|s^{N}\rangle$, $|1\rangle|g^{N}\rangle$ and $|1\rangle|s^{N}\rangle$ of the wave-vector as it evolves with time in the absence of any decay from the excited levels of the control and the target atoms. The final state obtained after measuring the control atom in $\frac{1}{\sqrt{2}}(|0\rangle+|1\rangle)$ state has a fidelity of 0.97 with respect to the GHZ state $|\phi\rangle =\frac{1}{\sqrt{2}}(|g^{N}\rangle+|s^{N}\rangle)$ for a target ensemble with N = 5 atoms and having the interaction strength $\tilde{\Delta}=500$.
\begin{figure}
\centering
    \includegraphics[width=0.49\textwidth]{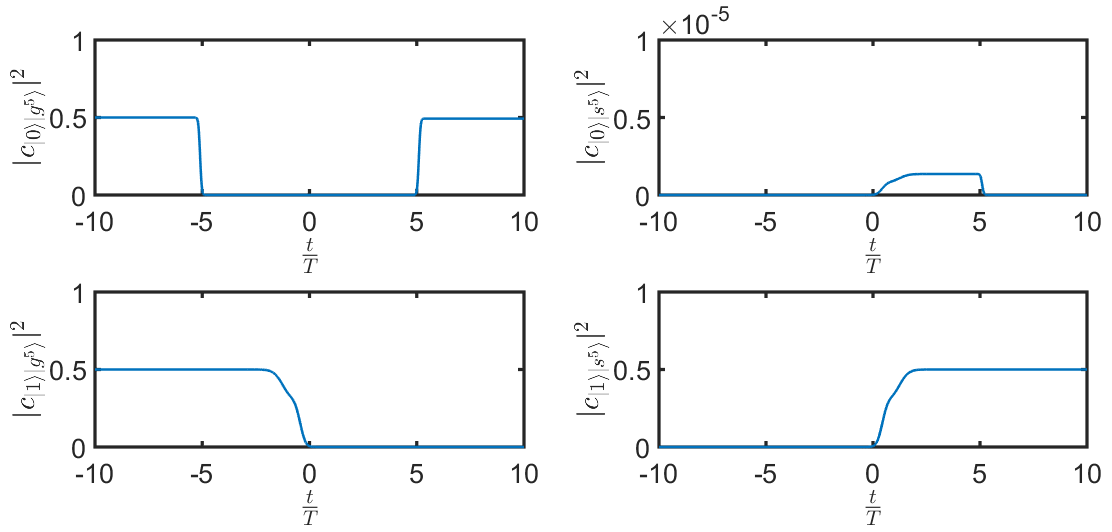}
\caption{Implementation of the protocol for N=5: Time evolution of the squared  co-efficients of $|0\rangle|g^{N}\rangle$, $|0\rangle|s^{N}\rangle$, $|1\rangle|g^{N}\rangle$ and $|1\rangle|s^{N}\rangle$ under the influence of the Hamiltonian in Eq. (\ref{eq:totH}) with the initial condition $\frac{1}{\sqrt{2}}(|0\rangle+|1\rangle)|g^{N}\rangle$. Chosen parameters: $\Omega_{c0}T=6.2$, $\delta_{R}T=0$, $T_{c} = 0.1T$,  $\Omega T=5$, $\delta T =0$, $\tau = 1.4T$, $\Delta T = 500$, $\Gamma_{r}T = \Gamma_{R}T = 0$, $\tau_{c}=\tau+4(T+T_{c})$.}\label{fig:5}
\end{figure}
Note that for this simulation, $T=1\mu s$, which means that the entire operation takes only about 15-20$\mu s$. Typical excited Rydberg level lifetimes for $n\gtrsim 60$ are of the order of 100 $\mu s$ \cite{Saffman05}. Since the current time of gate operation is much less compared to the excited level lifetime, we can improve the fidelity by increasing the value of $\tilde{\Delta}$ without necessarily exciting the Rydberg atoms to much higher levels by simply increasing the width of the STIRAP pulses. In Fig. (\ref{fig:6}), we plot the fidelity of the obtained final ensemble state wrt to the GHZ state $|\phi\rangle$ as a function of the interaction strength $\tilde{\Delta}$ for a target ensemble having 1 and 5 atoms. The fidelity for a single target atom is above 98\% for $\tilde{\Delta}$ of 100 or more. On the other hand the fidelity of the target ensemble is 98\% and higher for values of $\tilde{\Delta}=$ 600 and above.\\
\begin{figure}
\centering
    \includegraphics[width=0.4\textwidth]{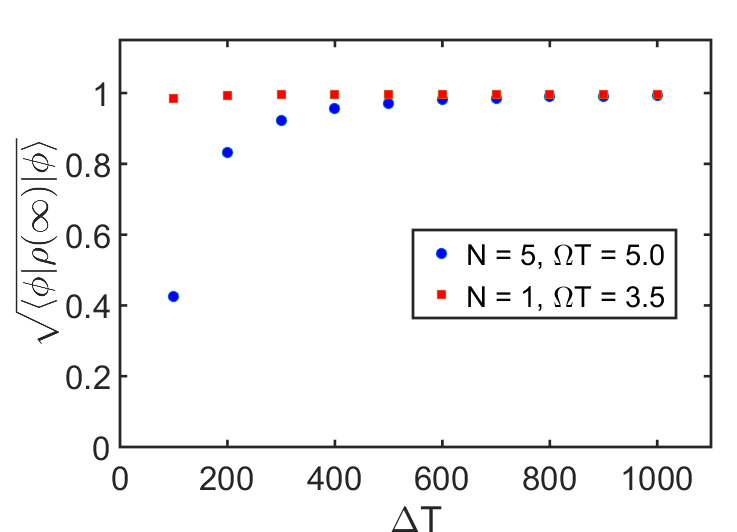}
\caption{The fidelity of the final ensemble state wrt to $|\phi\rangle$ for N= 1 and 5 as a function of the interaction strength $\Delta T$ with the initial condition $\frac{1}{\sqrt{2}}(|0\rangle+|1\rangle)|g^{N}\rangle$. Parameters used in the simulation: $\Omega_{c0}T=6.2$, $\delta_{R}T=0$, $T_{c} = 0.1T$, $\delta T =0$, $\tau = 1.4T$, $\Gamma_{r}T = \Gamma_{R}T = 0$, $\tau_{c}=\tau+4(T+T_{c})$}\label{fig:6}
\end{figure}
As we have already seen, the spontaneous emission from the excited levels of the target atoms do not affect this protocol as long as the adiabaticity conditions are satisfied. What about the spontaneous emission from the excited level of the control atom? In Fig. (\ref{fig:7}) we show the decrease in the fidelity of the final density matrix wrt the state $|\phi\rangle$ for the same initial conditions as above due to the decay from the $|R\rangle$ level. This plot shows the decay rate for the target ensemble having a single atom and 5 atoms with $\tilde{\Omega}=3.5$, $\tilde{\Delta}=200$ and $\tilde{\Omega}=5$, $\tilde{\Delta}=500$ respectively and $\tilde{\delta}=0$. As expected the rate of the decay is same for both the cases since the number of target atoms does not influence it. The fidelity is seen to drop to a value of 97\% from 99\% for a single atom target ensemble when the value of $\Gamma_{r}T$ increases to 0.01, whereas for the target ensemble with 5 atoms, the fidelity drops from 97\% to 95\%. \\
\begin{figure}
\centering
    \includegraphics[width=0.4\textwidth]{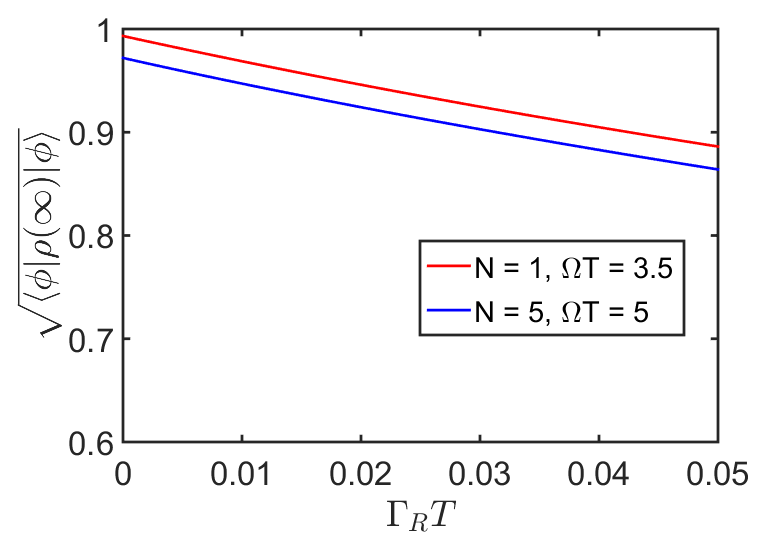}
\caption{ Fidelity of the target ensemble density matrix after measurement of the control atom in the superposition state measured  wrt to the state $|\phi\rangle$ with N = 1 and 5 for different values of $\Gamma_{R}T=\Gamma_{r}T$ numerically evaluated with the initial condition $\frac{1}{\sqrt{2}}(|0\rangle+|1\rangle)|g^{N}\rangle$. Parameters: $\Omega_{c0}T=6.2$, $\delta_{R}T=0$, $T_{c} = 0.1T$,   $\delta T =0$, $\tau = 1.4T$, $\Delta T = 200$ for N = 1, $\Delta T = 500$ for N = 5,  $\tau_{c}=\tau+4(T+T_{c})$}\label{fig:7}
\end{figure}

It is possible to compensate for the losses due to spontaneous emission from the control atom by exciting it to higher Rydberg levels. This would serve the dual purpose of providing longer excited level lifetimes as well as stronger Rydberg dipole interaction strength \cite{Saffman10}, which would in turn improve the overall fidelity of the protocol.

\section{\label{sec:level5}Conclusion}
In conclusion, we have presented here a protocol to create N particle GHZ state with a single control atom and an ensemble of N target atoms based on the principles of Rydberg dipole blockade and STIRAP. We have discussed the conditions under which adiabatic transfer of the target ensemble population from one ground state to the other is facilitated subject to the state of the control atom. The biggest advantage of this scheme is that it is not affected by the decay from the excited Rydberg levels of the target ensemble atoms as long as the conditions for adiabatic transfer are satisfied. Spontaneous emission from the excited Rydberg level of the control atom leads to decrease in the fidelity of the protocol. This can be controlled for by exciting the control Rydberg atom to higher principal quantum number.\\
\section*{Acknowledgement}
TG would like to thank Prof. Luming Duan, University of Michigan, for helpful discussions and his valuable guidance throughout this project.    
\bibliography{research2}
\end{document}